\documentclass[conference]{sig} %
\usepackage{graphicx}
\usepackage{xspace}
\newcommand{\BfPara}[1]{{\noindent {\bf #1.}}}
\newcommand{\etal}{{\em et al.}\xspace}

\usepackage{times}
\usepackage{color}
\usepackage{listings}
\usepackage{balance}
\begin{document}

\title{Rethinking Information Sharing for \\ Actionable Threat Intelligence\thanks{Approved for Public Release; Distribution Unlimited: 88ABW-2016-3801, dated 29 July 2016}}



\numberofauthors{4} 
%
\author{
%
%
\alignauthor Aziz Mohaisen\thanks{This work was done in part while the author was visiting the US Air Force Research Lab, in Rome, NY on the summer faculty fellowship program (summer 2016).}\\
       \affaddr{University at Buffalo}
       \email{mohaisen@buffalo.edu}
\alignauthor Omar Al-Ibrahim\thanks{This work was done while the author was a visiting scholar at the University at Buffalo.} \\
	\affaddr{University at Buffalo}
	 \email{omaralib@buffalo.edu}
\alignauthor Charles Kamhoua \\
       \affaddr{ Air Force research Lab}
       \email{charles.kamhoua.1@us.af.mil}
\and
\alignauthor Kevin Kwiat\\
       \affaddr{ Air Force research Lab}
       \email{kevin.kwiat@us.af.mil}
\alignauthor Laurent Njilla \\
       \affaddr{ Air Force research Lab}
       \email{laurent.njilla@us.af.mil}
}

\maketitle

\begin{abstract}
In the past decade, the information security and threat landscape has grown significantly making it difficult for a single defender to defend against all attacks at the same time. This called for introducing information sharing, a paradigm in which threat indicators are shared in a community of trust to facilitate defenses. Standards for representation, exchange, and consumption of indicators are proposed in the literature, although various issues are undermined. In this paper, we rethink information sharing for actionable intelligence, by highlighting various issues that deserve further exploration. We argue that information sharing can benefit from well-defined use models, threat models, well-understood risk by measurement and robust scoring, well-understood and preserved privacy and quality of indicators and robust mechanism to avoid free riding behavior of selfish agent. We call for using the differential nature of data and community structures for optimizing sharing.

\end{abstract}
\section{Introduction}
With the emergence of new information and communication technology platforms, such as cloud computing, mobile computing, social networks, and the Internet of Things (IoT), the security landscape has become more sophisticated in the past decade.  What used to be an unmotivated form of vandalism during the early days of the Internet has become a diverse ecosystem of cybercrime, where providers and consumers come together to achieve various end-goals and utilities. The persistence, complexity, size, and capabilities of today's adversaries are unbounded, and their threat does not only affect individuals or organizations, but also nations as a whole:  according to a recent study~\cite{study1}, direct and indirect costs due to security breaches have costed the global economy about \$491 billion in 2014 alone. 

\BfPara{The need for information sharing}
Defending against the unknown is a difficult task~\cite{BowersDGJORT12}. Accordingly, visibility into the behavior and capabilities of adversaries to form detection signatures is an essential first step towards containing and defending against them, and ultimately thwarting their harms~\cite{HongXWG15}. On the other hand, with the unprecedented complexity and size of the threat ecosystem, no single defender can defend against all attacks all the time. Even when facing attacks, defenders need to have the right skills to recognize them before performing defense efforts. With the skill gap on the rise, visibility into attacks and malicious actors becomes a challenge.  Thus, a coordinated solution based upon the collective knowledge of multiple defenders is required. In such a solution, multiple stakeholders share information about security incidents observed and collected from their security operations, with the hope that such information would be useful to other stakeholders in improving their security posture. 




\BfPara{Information Sharing} Recently, information sharing as a concept has emerged as a plausible solution to addressing the aforementioned problems. Threat information sharing is utilized for efficiently and effectively defending against emerging threats. One even went as far as to say that ``threat intelligence sharing is the only way to combat the growing skills gap''~\cite{skillgap}. In practice, information sharing is used to communicate operational security experience between a set of participants in a sharing system with the hope that sharing would 1) enable participants to defend their systems against ongoing attacks, and 2) improve their defense posture by proactively addressing possible attacks even before they target them.  


\BfPara{Initiatives of information sharing} Information sharing is not a theoretical idea, and there has been a lot of work in the past on defining tools for representing information, or mechanisms for exchanging such information between information sharing participants in sharing communities. Information sharing also has been embraced by various communities, and leaders in such community have created their own sharing exchange points, where participants could deliver and retrieve the shared raw data and annotated data (intelligence) from other participants using standard application program interfaces (APIs): for example, Facebook has created ThreatExchange~\cite{threatexchange}, and Verisign has created the IntelGraph, among others. Such initiatives are not limited to the private sector: the public sector initiated information sharing in the US Department of Homeland Security's (DHS) Critical Infrastructure Cyber Information Sharing and Collaboration (CISCP) program~\cite{ciscp}, which aims to facilitate sharing of threat indicators between the private and public sectors and vice versa.

\BfPara{Risks of sharing}
The risk of not sharing information is clear, which can be seen in more and more security breaches using the same attack vectors and capabilities, and re-examining the same vulnerabilities. Despite the various benefits of information sharing for security, even within a limited community of participants, shared information without proper restrictions, however, may leak a significant amount of information about the participants and their operation context. For example, information shared for the good and security of the participating community can be also used by the adversary to learn about vulnerabilities of those participants. Also, the same vulnerabilities can be used to test their applicability on other systems: with the lag in patching vulnerabilities, the adversary will be able to utilize such information for attacking other unpatched systems.  

This risk of sharing can be mitigated in information sharing by limiting the sharing community to highly trusted participants, and informing potentially subjected participants with the risk ahead of time, and certainly before sharing such information with a broader community. However, limiting sharing to a highly trusted community of participants in general security applications, while reducing risk of information exposure to unintended parties---including adversaries---may have an equally damaging consequence: the security of today's systems with the presence of multiple stakeholders exhibit the end-to-end principle of design characterized by fate-sharing. For example, an unpatched system under the control of uninformed player simply because that player is not trusted (enough) can be used to attack other systems under the control of the highly trusted participants.  Classification of participants' risk in sharing communities is provided in the literature, despite the lack clarity on how such risk should be assessed~\cite{framework}\footnote{Indeed, the work in~\cite{framework} while being seminal in implementing various ideas to realize an information sharing system, it does not address how such risk should be assessed, and how audit (for both security and privacy) should be performed. It rather leaves to users to decide which levels of risk they should assign to themselves}. 

Another scenario of sharing where negative consequences may arise is privacy of individuals, and how sharing may affect civil liberties~\cite{nist2}.  The sharing of public data that does not in of and itself identify individuals would serve the goal of information sharing without any side effects on privacy. However, it is believed that privacy does not often go along well with security: to be able to attribute attacks and perform various security analyses, context information should be present along with the threat indicators for further inferences that would serve security. For example, along with an end point (i.e., hostname, or IP address) an incident indicator typically shared would include e-mail addresses, URLs, etc., from which intelligence is gathered, and security risk is assessed.  

Privacy risks due to sharing are arguably mitigated by a minimalistic approach, where only a limited amount of data is {\em collected and shared}~\cite{framework}. However, whether such a minimalistic approach is being implemented in today's sharing paradigms or not is unclear. Furthermore, such an approach goes against security utilities. We conjecture (with confidence based on various plausible applications) that the additional context information of the threat information shared is often times as important as the indicators themselves. To this end, a new approach to thinking privacy is required beyond simple minimization. Such a technique could perhaps utilize techniques for safeguarding  all necessary information to improve the security posture of a defender, while ensuring the privacy of users, and confidentiality of shared information.

Related to the community of trust problem above and perceived risk of over sharing (whether it is for security or privacy) is the problem of ``free-riding''. As a result of the perceived risk of sharing, some actors might be actually joining communities of sharing, although not sharing sufficient information to the community that others can benefit from it~\cite{KamhouaMTKHS15,KamhouaRMK15}. When a community member joins and shares information there is always the risk of the shared information (e.g., about a vulnerability) being leaked to the public (or even worse, to the adversary), resulting in both monetary and reputation loses. Such scenario leads to that some actors might not truthfully share information due to their own self-interests. While recent works have been focused on addressing problem in a theoretical framework~\cite{ToshSMKK15a,ToshSKKM15b}, assuming the level of participation as an indicator of contribution in information sharing, there is no work that extends beyond that to account for quality of indicators. For example, an actor that contributes stale indicators, indicators that are not timely to be utilized operationally, while not considered a free-rider in the typical sense, is not contributing sufficiently and meaningfully to the missing of information sharing.


Believing in their beneficial aspects, the goal of this study is to shed light on various issues associated with information sharing, including understanding community structures, use and adversary models, privacy issues and quality of indicators for detecting free riding in  information sharing for actionable threat intelligence. With standard sharing formats being widely advocated as the next step towards effective sharing, we identify the need for understanding privacy and risk. To understand this risk in context, we identify a plausible sharing scenario for which we define the adversary models associated with both internal or external adversaries. We introduce to the analysis the various sharing paradigms under them. By identifying the need for security, we advocate an approach that combines various aspects of design techniques that exploit the differential nature of data and community structure. Finally, we identify quality of indicators as an important direction, suggest various directions to assessing quality, and call the research community to further the suggested directions. 

\BfPara{Organization} The rest of this paper is  organized as follows. In section~\ref{sec:rethinking}, we provide our broad vision of various directions for rethinking sharing towards actionable threat intelligence. In section~\ref{sec:privacy} we ellaborate on one of directions, namely privacy. In section~\ref{sec:quality} we elaborate on another issue, namely, quality of indicators. Concluding remarks are in section~\ref{sec:conclusion}.

\section{Rethinking Sharing}\label{sec:rethinking}
Realizing actionable intelligence by striking a balance between utility of the information sharing systems and other requirements, including privacy, security, and complexity of the sharing system, is a non-trivial task. In the following, we offer to rethink sharing by touching on various fundamental issues and building blocks in typical sharing systems. We identify the following issues as rich areas that require further research and exploration, and offer various directions associated with each of those issues in the subsequent sections. We offer to understand use models (\textsection\ref{sec:use}), sharing communities (\textsection\ref{sec:communities}), adversaries in sharing paradigms, including both outsider and insider adversaries (\textsection\ref{sec:adversaries}), quantifying understanding privacy in information sharing, towards measurement

\subsection{Defining the Use Model}\label{sec:use}
Information sharing is inseparable from its use model and scenario. Thus, understanding the various technical details of the use model of information sharing tools and paradigms is essential to understanding various issues, including security, privacy, and functional issues. We offer to touch on various scenarios of use and issues associated with in the following. 

%

We classify the use models of information sharing for threat intelligence into various types based on various classification criteria, as follows: 
\begin{itemize}
\item {\sf Structure:} Based on the {\em structure} and format of the shared information, we classify information sharing tools into structured (standard) and unstructured sharing models. 
\item {\sf Centralization:} Based on whether a centralized sharing entity (repository) exists or not, we classify such models into centralized and decentralized systems.
\item{\sf Scope and function:} Information sharing tools can be also classified based on their scope and function. While it is difficult to enumerate such scopes and functions for unstructure sharing systems, structured systems that use standards are classified into enumeration, scoring, languages, and transport mechanisms. More details are provided in~\textsection\ref{sec:privacy}.
\end{itemize}

\BfPara{Unstructured Sharing}
The end goal of information sharing is to realize a secure cyberspace by exchanging operational security experience across multiple players in a sharing community. Whether data used in the information sharing paradigm is structured or not is irrelevant to the main goal of information sharing. Traditionally, threat information concerning incidents has been collected and shared as unstructured data, and exchanged via generic communication tools and services, including electronic mail, or file transfer services. Today, and despite the rise of structure via standardized formats and sharing schemas, proprietary formats are widely used in by vendors in security  market, making interoperation between structures hard to achieve. While it is easier to understand structured schemas, where various attributes are indicated, understanding the privacy of sharing when using unstructured formats is not possible. To this end, in the rest of this work we focus on structured sharing format, although we believe that unstructured sharing also may have various privacy risks that should be studied and addressed based on actual  assessments. 

\BfPara{Sharing Using Standard Formats}
For efficient use of shared information in an automated manner, it is desirable to share information in a standard and structured format. For that, there has been a lot of work in the literature on understanding use scenarios, and developing the relevant schemas of structured formats for information sharing.  By understanding the type of data in such information sharing formats, it would not only be possible to understand the capabilities embodied in the various sharing formats, but also to understand the privacy risks in the abstract, and possibly develop technical solutions to address it. Examples of such sharing paradigms include CVE, CCE, CWE, etc. CyBox, etc. More on those schemas and formats are in \textsection\ref{sec:schemas}.

\subsection{Defining Communities}\label{sec:communities}
Sharing is defined around ``communities of trust'', which are the structure in which (potentially) threat information is shared to reach a common goal of strengthening the security posture of various participants in the community. Sharing today is defined based on the nature of the participants (whether they are public or private sector participants) into private-private, public-private, and public-public. An example of the private-private sector information sharing communities include participants in the likes of ThreatExchange, or IntelGraph, while an example of the public-private partnernships include DHS's CISCP~\cite{ciscp}. 

On the one hand, various of those communities are vetted carefully to ensure that the information being shared between the various participants in the sharing system is safeguarded and not used to attack any participant in the system. On the other hand, circumstantial evidence (or even conclusive evidence~\cite{ryan2012use}) has shown that information being shared in the sharing system could potentially be used as an attack vector against another participant in the system. Understanding the make up of the sharing community is perhaps a first step to account for such risk. 

\BfPara{Redefining communities} Redefining communities structure by relaxing the meaning and assumptions of ``trust'' in a way that would allow for a greater participation of players in a sharing system results, thus potentially resulting for improved defenses and security awareness by a larger number of participants, would potentially result in a higher risk of sharing. Such risk is not only seen in increasing the security attack surface, but potentially in disincentivizing major community members from meaningful and sharing of quality information that could result in actionable intelligence. Understanding how relaxing the definition of the community would affect both utility of the sharing system and the risk of sharing is to be considered further in light of actual and measurable risk.  

\BfPara{Privacy-based community definition} So far, communities have been defined for their trustworthiness with respect to their risk awareness, or for utilizing the various tools and paradigms of information sharing, but not understood with respect to privacy. Thus, we believe it is a worthwhile to incorporate privacy as a metric (along with other metrics of risk or in isolation) as a criteria for defining communities. Furthermore, technical solutions that take into account a clear definition of privacy-awareness and its presence (or lack) in a certain community (or players in a community) could be further optimized to suite the underlying assumptions of such community.



\subsection{Redefining Adversaries}\label{sec:adversaries}
Security and privacy of communication and computation protocols are often analyzed under various settings of adversaries. Adversaries are characterized by capabilities under which security and privacy definitions are formalized, and security and privacy guarantees (in light of a formally defined advantage of the adversary) are outlined. With the complexity and involved nature of information sharing paradigms, and the end-goal that they try to achieve, we argue that both insider and outsider (external) adversaries are relevant to studying the information sharing in the field.  In the following, we elaborate on both forms of threat, and open directions to address in order to realize a formally-backed exchange. 

\BfPara{External adversary} Outsider adversaries in the context of information sharing are defined broadly as adversaries who are not part of the system or protocol being analyzed, and they include various forms of actors, ranging simply from a passive eavesdropper~\cite{ref3,goldwasser1997multi} or honest-but-curious~\cite{kissner2005privacy} to the more advanced active adversary--an adversary that could potentially interfere with communication or manipulate computations in order to affect the security of the system, or breach the privacy of a participant. This adversary can be a single malicious actor, or multiple of them. The main qualifier of this adversary, however, is that it is not included in the set of participants of the system. 

Examples of instances of such adversary include simple observers on the communication channel between participants in the information sharing system, with their risks being mitigated by the various in-place cryptographic techniques. Another example of the observer could be a publicly shared infrastructure, like cloud, where the cloud provider may have a great incentive not to act maliciously, but would be interested in knowing some details about the information being shared and hosted in the cloud. While auditing and strict policies are one direction to tame this adversary, relinquishing trust and enforcing a stronger form of audit---perhaps by utilizing cryptographic approaches, is yet another method. We elaborate on such methods in the subsequent sections.  

The same example above of cloud could be also viewed as a totally untrusted, and potentially malicious, thus being an instance of the malicious adversary. Such state of being malicious could be a property of the cloud itself; i.e., the cloud provider is untrusted, or due to other externalities, e.g., the cloud is being compromised by an outsider through, for example, a malware campaign. The way that such adversary is realized is irrelevant to understanding the privacy of the various sharing paradigms, although the capabilities of such adversary are.

\BfPara{Insider adversary} Motivated by the various risks that potentially could be the result of misuse of the information shared an information sharing system~\cite{ryan2012use}, another adversary model that needs to be formalized is the insider adversary. Whereas typical threats in various systems include the external adversary highlighted above, the nature of information sharing systems highlight that insider adversaries are real risks. Such adversaries could be in multiple forms, and stem out of various system and operation realities. For example, such adversary could be another participant in the information sharing system acting maliciously to reach a certain objective, or an individual acting on behalf of a participant in the system. 

Understanding how information sharing is prone to such class of adversaries is necessary to enable sharing. Furthermore, such adversary could perhaps be studied well under other notions of risk associated with information sharing and definition of communities of trust, their risk and privacy awareness. 

\subsection{Measurements}\label{sec:measurements}
One may argue that the problem at hand is not any different from any other privacy problem due to data exposure, thus thinking of the privacy issues in information sharing for threat intelligence in the abstract is meaningful and the way this problem should be addressed given the large number of use scenarios. 

We argue that while thinking of this problem in the abstract is worthwhile, also approaching the problem with technical solutions that stem from the actual size and shape of privacy exposure in the various information sharing paradigms is equally important. A first step towards understanding the actual size and shape of exposure is facilitated by an actual quantification of exposure in real data. However, one cannot quantify what he cannot measure, thus measuring data exposure in the various  sharing paradigms, under the various settings of threat models or in isolation, is necessary and important for understanding the problem in practice. In particular, measurements would give analytical and abstract studies context that highlight the actual meaning of findings related to indicators, privacy, and risk. 

Measuring privacy leakage in the various paradigms of sharing and under various use models is not an easy task. We argue that privacy cannot be understood in the abstract, and without a clear context of the sharing~\cite{nissenbaum2009privacy}. Even worse, what constitute a privacy concern to one individual might not be of value to another individual in the same context. Thus, a first step to measuring privacy leakage in information sharing is to formalize what we mean by privacy, what are the private attributes that should be treated with care and hidden from adversaries and other (potentially honest-but-curious) participants, and how sensitive (with respect to their privacy value) alone or when associated with other data about the subject.

\subsection{Quality and Privacy}\label{sec:qp}
Quality of the indicators and privacy are at odds: in order to provide the highest accuracy in security operations, access to raw and highlight annotated indicators that can be of use for actionable intelligence is necessary. On the other hand, having such raw indicators without any sanitization or masking of any of their contents could potentially leak the privacy of entities associated, or reveal sensitive information about the operation context where they are collected, directly or indirectly. To this end, another direction to pursue is by answering the following question: {\bf How much quality of indicators should be given up to satisfy various privacy notions and guarantees}. 

This question is not easy to answer: there are various competing and varying notions of privacy, and  systematically and formally analyzing and modeling how they  are met (or violated) at various levels of exposure of indicators. Before even approaching this question, it would be necessary to formalize metrics for evaluating the quality of the indicators. 

\section{Understanding Risk in Sharing}\label{sec:privacy}
There is a clear risk of sharing, whether it is privacy or security. Understanding such risk is the first step towards providing practical solutions to the various aspects of risk. In the following, we elaborate on a road-map for understanding risk in information sharing, mainly emphasizing privacy risks. In \textsection\ref{sec:schemas} we review the various sharing schemes. In \textsection\ref{sec:risk} we highlight risks of information sharing through various measurements and examples from {\em anonymized} sharing datasets. In \textsection\ref{sec:assessment} we argue for a privacy leakage assessment design that takes into account the various issues raised on the risk of the sharing paradigms. In \textsection\ref{sec:designs} we advocate architectural design that takes into account privacy and community structure as a design principle. 
\subsection{An Overview of Sharing Standards}\label{sec:schemas}

As noted previously, there are various standards for information sharing that are used by government and industry to automate and structure the exchange of information within an organization and between autonomous systems and organizations.  We can classify these sharing standards into four main categories:

\BfPara{Enumerations} Standardized enumerations of platforms, configurations, software weaknesses, and attacks. Examples include Common Configuration Enumeration (CCE), Common Weakness Enumeration (CWE), and Common Vulnerabilities and Exposures (CVE). A listing of such enumeration techniques is shown in Table~\ref{tab:enumeration}.

\begin{table*}
\begin{center}
\caption{Enumeration Standards}\label{tab:enumeration}
{\scriptsize
\begin{tabular}{|p{7cm}|p{9cm}|}\hline
Name	&	Description	\\ \hline
\underline{C}ommon \underline{V}ulnerability \underline{E}xposure	&	Standard identifiers for publicly-disclosed cybersecurity vulnerabilities.	\\ \hline
\underline{C}ommon \underline{W}eakness \underline{E}numeration	&	Standard identifiers for software weaknesses or flaws. 	\\ \hline
\underline{C}ommon \underline{A}ttack \underline{P}attern \underline{E}numeration and \underline{C}lassification	&	Enumeration of cyber-attack techniques.	\\ \hline
\underline{C}ommon \underline{C}onfiguration \underline{E}numeration	&	Enumeration of configurations covering various platforms and controls ref.	\\ \hline
\underline{C}ommon \underline{P}latform \underline{E}numeration	&	Standard identifiers for platforms, operating systems and applications.	\\ \hline
\end{tabular}
}
\end{center}
\end{table*}

\BfPara{Scoring systems} Standards to assess the severity of computer system-related issues and assigning scores to each one, allowing responders to prioritize remediation tasks. Common standards that fit this category include Common Vulnerability Scoring System(CVSS) and Common Weakness Scoring System(CWSS). A listing of such scoring systems is shown in Table~\ref{tab:score}.

\begin{table*}[htb]
\begin{center}
\caption{Scoring System Standards}\label{tab:score}
{\scriptsize
\begin{tabular}{|p{7cm}|p{9cm}|}\hline
Name	&	Description	\\ \hline
Common Vulnerability Scoring System (CVSS)	&	Standard rating for calculating a score on severity of vulnerabilities.	\\ \hline
Common Weakness Scoring System (CWSS)	&	Risk rating of software vulnerabilities considering access complexity. 	\\ \hline

\end{tabular}
}
\end{center}
\end{table*}

\BfPara{Languages} Those sharing standards are intended for encoding high-fidelity information about systems in a manner that facilitates parsing this information in software tools and converting them to human-readable formats. This includes formats like Incident Object Description Exchange Format (IODEF) and Open Vulnerability and Assessment Language (OVAL). A listing of such standards is shown in Table~\ref{tab:lang}. 

\begin{table*}[htb]
\begin{center}
\caption{Language Standards}\label{tab:lang}
{\scriptsize
\begin{tabular}{|p{7cm}|p{9cm}|}\hline
Name	&	Description	\\ \hline
Malware Attribute Enumeration and Characterization (MAEC) 	&	Language to describe malware characteristics, behavior, and actions.	\\ \hline
Open Vulnerability and Assessment Language (OVAL)	&	Language to express machine state, writing assessment tests and reporting results. 	\\ \hline
Incident Object Description Exchange Format (IODEF) & Format to define computer and network security-related incidents. \\ \hline
Extensible Configuration Checklist Description Format (XCCDF) & Language for writing security checklists, benchmarks and other related documents. \\ \hline
Structured Threat Information Exchange (STIX) & Format to express information about cyber-security threats. \\ \hline
\end{tabular}
}
\end{center}
\end{table*}

\BfPara{Transport} Those standards represent Inter-network communication formats to facilitate exchange of information between hosts. Standards such as Real-time Inter-network Defense (RID) and Trusted Automated eXchange of Indicator Information (TAXII) fit this category.
In the following, we elaborate on the different category of standards and how they are used to automate information sharing within organizations. A listing of such standards is shown in Table~\ref{tab:transport}. 

\begin{table*}
\begin{center}
\caption{Transport Standards}\label{tab:transport}
{\scriptsize
\begin{tabular}{|p{7cm}|p{9cm}|}\hline
Name	&	Description	\\ \hline
Real-time Inter-network Defense (RID) 	&	Transport standard that facilitates sharing of incident-handling data, typically  stored in IODEF-formatted documents.	\\ \hline
Trusted Automated eXchange of Indicator Information (TAXII) 	&	Standard that defines protocols and messages to exchange cyber-threat information. 	\\ \hline
Simple Object Access Protocol (SOAP) & Messaging framework typically used in implementation of web services. \\ \hline
Reputation Services (Repute, DKIM) & Standards that define query methods for reputation data services. \\ \hline
\end{tabular}
}
\end{center}
\end{table*}

\lstset{ %
language=xml,                
basicstyle=\fontsize{3}{4}\ttfamily\footnotesize,       
numbers=left,                   
numberstyle=\footnotesize,      
stepnumber=1,                   
numbersep=5pt,                  
backgroundcolor=\color{white},  
showspaces=false,               
showstringspaces=false,         
showtabs=false,                 
frame=single,           
tabsize=2,          
captionpos=b,           
breaklines=true,        
breakatwhitespace=false,    
escapeinside={\%*}{*)}          
}

\newcommand{\hlcyan}[1]{{\sethlcolor{cyan}\hl{#1}}}
\newcommand{\hlred}[1]{{\sethlcolor{red}\hl{#1}}}

\subsection{A Privacy Risk in Standards}\label{sec:risk}

In this section, we highlight the various risks associated with information sharing. For that, {\em anonymized examples depicted from sharing operations utilizing the standard schemas for information sharing}. For illustration, we label the leaking fields with different colors depending on class of data being exposed, specifically, we designate red color for PII fields, light blue for non-PII sensitive fields (e.g., related to business context), and yellow for inference-leaking fields.\\[1pt]

\newcommand{\cbox}[1]{\raisebox{\depth}{\fcolorbox{black}{#1}{\null}}}

\textbf{Legend:}\\
\begin{center}
\begin{frame}

    \cbox{yellow} Inference \quad
    \cbox{cyan} Sensitive \quad
    \cbox{red} PII 
\end{frame}
\end{center}
\vspace{2mm}
In the following, we highlight such risk through various examples obtained from real shared information. 

\begin{figure}[htb]
\centering
\includegraphics[width=0.47\textwidth]{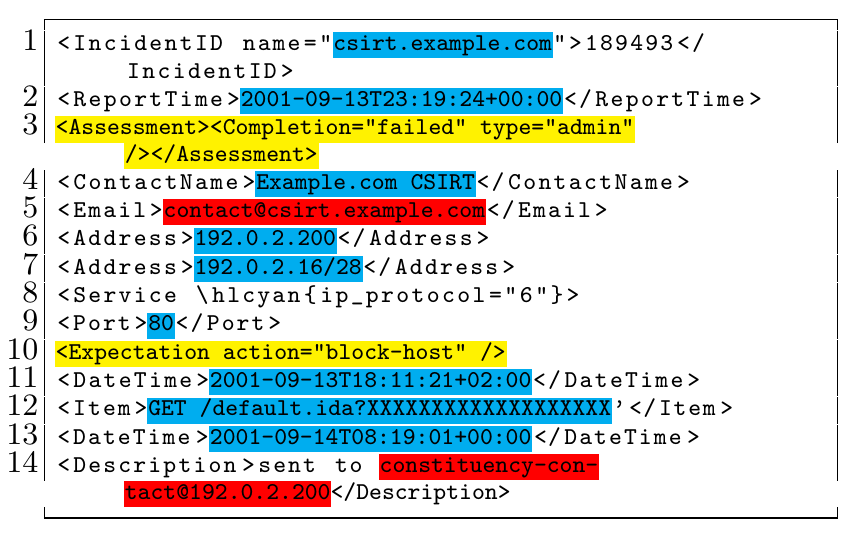}
\caption{Annotated private, sensitive, and confidential information (inference) information in IODEF.}\label{fig:iodef1}
\end{figure}

\BfPara{IODEF worm report}
An example of a CSIRT reporting an instance of the Code Red worm, encoded in IODEF, is depicted in Figure \ref{fig:iodef1} (notice that a substantial part of the document is omitted, and only essential information is shown for demonstration). As shown, the document contains contact information (name, registry handle, email) for the constituent responsible for the incident report. This type of information may become personally identifiable in the case when contact information of a particular individual is used. The document also includes other fields that are less sensitive. This includes reporting time, record datetime, IP addresses of the node or network that were targeted in the attack, as well as the targeted service port number.

In this example, the Code Red worm attempted to target the HTTP port for a host machine with an IP address of \emph{192.0.2.200}. The raw HTTP request sent by the worm is also captured in the reporting. The worm intended to fiddle with the web server and the request was presumably an attempt for a buffer overflow attack in order to escalate to administrative privileges. Consequently, if the worm was successful in gaining access to the machine the information captured from the raw HTTP request may become highly sensitive. However, we know from the ``assessment'' field in the document that it was a failed attempt.
\begin{figure}[htb]
\centering
\includegraphics[width=0.47\textwidth]{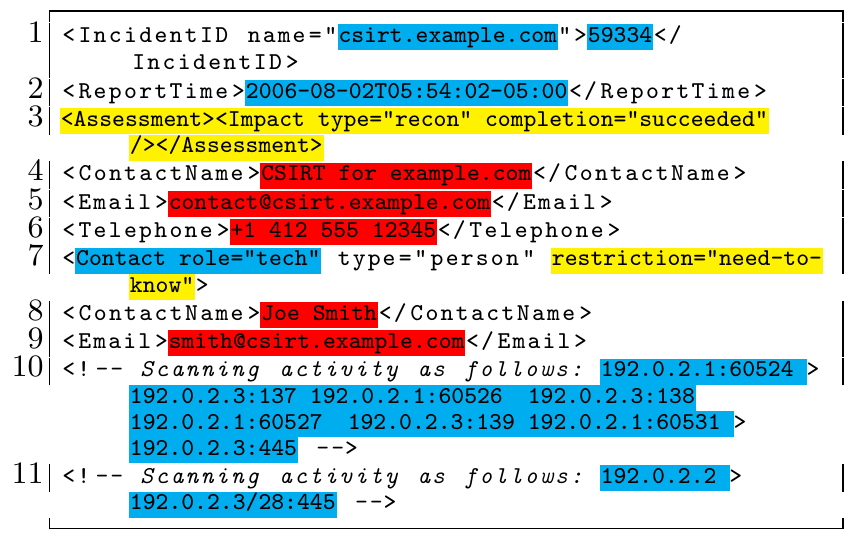}
\caption{Annotated private, sensitive, and confidential information (inference) information in IODEF}\label{fig:iodef2}
\end{figure}
Another example, with some information masked, based on the same standard is shown in Figure~\ref{fig:iodef2}.

\begin{figure}[htb]
\includegraphics[width=0.5\textwidth]{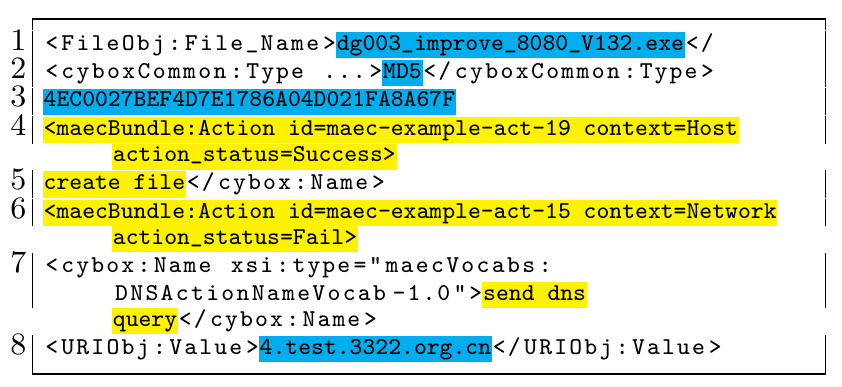}
\caption{MAEC for dynamic triage}\label{fig:maec1}
\end{figure}

\BfPara{MAEC package dynamic triage} An example to demonstrates how a package using the MAEC standard can be used to capture multiple dynamic analysis tool outputs for a malware instance is shown in Figure~\ref{fig:maec1}---shortened for summary only. It builds upon static triage example that shows the actions and details of the process tree associated with the instance. As depicted, the packaged output has few fields that may be considered sensitive, such as the domain name of the command \& control server (\emph{reallybadguy.com}; only for illustration). Exposing the domain name server to entities outside of the trusted community may inform the attacker about the detection of their malware instance, and thus link the malware reported by the subject with a campaign.
 
In addition to exposing the domain name field, the output also includes a field about bundle actions that are associated with the malware and the status of these actions.  In this example, the malware successfully created a file on the host filesystem but failed to resolve the DNS for the command control server.

\begin{figure}[htb]
\centering
\includegraphics[width=0.45\textwidth]{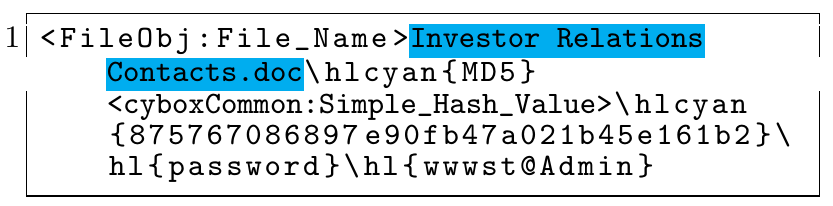}
\caption{Information leakage when using MAEC for information sharing for configuration parameters.}\label{fig:maec}
\end{figure}

\BfPara{MAEC package configuration parameter} Another example using MAEC, which demonstrates how to specify configuration parameters related to a malware instance characterized by a malware subject, is shown in Figure~\ref{fig:maec}. The configuration parameters are captured for tools such as remote access tools or reverse shell utilities and may include mutual exclusions, passwords, or IDs, for example. As shown below, a password value is specified as a configuration parameter. This may not be highly critical for malware samples that are password-protected, but it can be if the credentials provided in these fields are for accessing real sensitive documents or remote services.

Other countless examples that demonstrate various levels of risk to information that falls under one or more of the categories above exist. By showing those examples above, we hope to trigger interest in the community of pursuing research on understanding the level of leakage (and its context) for various of those sharing schemes in various application contexts. 


\subsection{Privacy Leakage Assessment}\label{sec:assessment}
With a clear understanding of what constitute attributes that would result in privacy violation, the presence or absence of such attributes in one instance of sharing could be used to assess privacy leakage in total. Informally speaking, we define a {\bf privacy leakage metric}, a single number associated with instances of standards when fully utilized to quantitively analyze the existing (potentially) private information in them. This metric can then be used to obtain numbers for each field in the schema of the standard, that could be aggregated to reflect a single score on the privacy level of other words. 

An example scoring system that assigns real values to various pieces of information in a sharing standard is as follows (just an example).\footnote{The basic question we pose is the creation of such privacy-oriented scoring system rather than realizing an exact numbering that would suite all applications: various information sharing applications weigh privacy differently, and may adjust individual scores differently.}  
\begin{itemize}
\item {\sf Score 0}: Public data or non-leaking
\item {\sf Score 1}: Inferential data
\item {\sf Score 2}: Sensitive data
\item {\sf Score 4}: PII data
\end{itemize}

The aggregate score is simply a summation of the scores for each field in the data schema of the standard. This is, let the schema of a standard have attributes $a_1, a_2, \dots a_n$. Based on the scoring system above, we generate scores $s_1, s_2, \dots, s_n$, which correspond to the various attributes. Using scoring system, we then calculate a single score for the standard, namely $s = \sum_{i=1}^n s_i$. 

Notice that the weights on the scoring metric can be modified depending on the level of emphasis on each class of data.  In this example, one point is given for an inferential field, two points for a sensitive field, and four points for a PII field. Since PII is protected through regulations, we give more weight for a PII field to be twice as significant as sensitive field and four times as significant as inferential field. Interpreting the score is trivial; the standard is considered highly privacy-leaking when the score number is large and more privacy-friendly when the number is small. 

Table~\ref{tab:leak} shows privacy leaking fields in schemas of various information sharing standards using and following a similar analysis to the one in the previous section. As can be seen, out of the four classes of standards the languages class have leaked the most personally identifiable information fields, particularly CyBox with a score of 56, STIX with a score of 36 and XCCDF with a score of 38. We notice that those standards embody various standards, and are inclusive of a large schema (with a large number of fields and attributes that cover multiple applications). Some standards have leaked fields that are considered sensitive but not PII, such as timestamps, host attributes like IP addresses, and organizational information. Other standards, such as enumeration standards, have mostly inference leaks related to vulnerability indices and platform identification numbers, which could potentially utilized and misused by an adversary.

Notice also that we do not advocate a specific scoring system for the risk assessment, since such scoring is context dependent. For example, an organization that views confidential information, information to do with business-related matters, might score confidential and inferential information higher than PII, since PII is not important to their security operation.

\begin{table*}[htb]
\begin{center}
\caption{Privacy leaking fields in schemas of various information sharing standards and example risk assessment using the indicated scores for the various leaked attributes. Scores are for illustration only.}\label{tab:leak}
{\tiny
\begin{tabular}{|p{1.5cm}|p{1.5cm}|p{3.5cm}|p{4cm}|p{3cm}|c|}
\hline
Standard Category & Standard & \multicolumn{4}{c|}{Privacy Leak}	\\ \hline
 & & PII (4) & Sensitive (2) & Inference (1) & Score \\ \cline{2-6}
Enumeration & CVE & & & CVE-ID & 1	\\ \cline{2-6}
 &CWE & & &CWE-ID & 1 \\ \cline{2-6}
 &CAPEC & & Submission:Source, Organization, Date & Relationship:ViewID, TargetForm, Nature, TargetID & 10 \\ \cline{2-6}
 &CCE & & cce:modified\_reference & cce:cce\_id, cce:platform & 4 \\ \cline{2-6}
 &CPE & & cpe:title & cpe:platform\_id & 3 \\ \hline
Scoring Systems & CVSS & & & &0 \\ \cline{2-6}
& CWSS & & & & 0 \\ \hline
Languages & OVAL & contributor & timestamp, submitted:date, status\_change, affected:family, platform, title, description & definition, reference & 20 \\ \cline{2-6}
& XCCDF & Benchmark:metadata, test:identity & cpe2:platform-specification, platform, status, test:organization, test:profile, test:target, test:target-address, test:target-facts, test:target-id-ref, test:start-time, test:end-time, test:fact & affected:family, platform, benchmarkIdType, resolved, test:authenticated, test:priviledged & 38 \\ \cline{2-6}
& MAEC & CommentType:author & ArtifactObj:Raw\_Artifact, maecPackage:Configuration\_Parameter, maecPackage:Name, maecPackage:Value, maecBundle:Collections.timestamp, AnalysisType:start\_datetime, AnalysisType:complete\_datetime, AnalysisType:complete\_datetime, AnalysisType:lastupdate\_datetime, AnalysisType:Comments, CommentType:timestamp & maecBundle:Action, maecBundle:CVE & 26 \\ \cline{2-6}
& CEE & & time, host, dst, ipv4, ipv6, src, port & status & 15 \\ \cline{2-6}
& IODEF & Contact, IncidentSource & DetectTime, StartTime, EndTime, ReportTime & Assessment, IncidentID, AlternativeID & 19 \\ \cline{2-6}
& STIX & stixCommon:Identity, stixCiqIdentity:Specification, xnl:PersonName, stixCommon:Name, xpil:Address, xpil:ElectronicAddressIdentifier, xpil:ContactNumber & timestamp, xpil:OrganizationInfo, xnl:OrganisationName, xpil:Nationalities/xpil:Country/xal:NameElement & & 36 \\ \cline{2-6}
& Cybox &  EmailMessageObj:Recipient, EmailMessageObj:From, AddrObj:Address\_Value, EmailMessageObj:Raw\_Header, Contributors,  ContributorType: Role, Name, Email, Phone  & HTTPSessionObj:Value, URIObj:Value, PortObj:Port\_Value, ArtifactObj:Raw\_Artifact, EmailMessageObj:Date, X509CertificateObj:Subject, X509CertificateObj:Issuer, TimeType: Start\_Time, End\_Time, Produced\_Time, Received\_Time, Observation\_Location, Observable\_Location, ContributorType:Organization, Date, Contribution\_Location & & 65 \\ \hline
\end{tabular}
}
\end{center}
\end{table*}

\subsection{Architectural Solutions for Privacy}\label{sec:designs}
A first step towards ensuring privacy is understanding the risk highlighted earlier through the actual sharing paradigms. Using a concrete notion of privacy, it would be then required to provide a technical solutions to meet such privacy notion, while enabling queries on the data shared using the various standards. In the following, we advocate architectural design that takes privacy and community structure into account for actionable intelligence through sharing. We start by reviewing the various notions of privacy, and then highlight the design space that could be exploited to ensure privacy. 

\subsubsection{Privacy Notions}\label{sec:notions}
Over the years, there has been various attempts in the literature to define the notion of the privacy, and provide techniques to ensure it. In the following, we review three notions that are of relevance, and advancing them could potential aid addressing the problem of privacy in the domain of information sharing: the $k$-anonymity, the $l$-diversity, the differential privacy
 
\BfPara{$k$-anonymity} Sweeney \cite{sweeney2002k} formulated the concept of \emph{k-anonymity} as an attempt to solve the problem of anonymizing person-specific field-structured data with formal guarantees while still producing useful data. A dataset is said to have the \emph{k}-anonymity property if the information for each person in the dataset cannot be distinguished from at least \emph{k-1} individuals.  Two common methods are available for achieving \emph{k}-anonymity, namely supression, where certain values of the dataset attributes are left blank, and generalization, where individual values of attributes are replaced by with a broad category, typically a range of values. Meyerson and Williams \cite{meyerson2004complexity} demonstrated that optimal \emph{k}-anonymity is an NP-hard problem, however heuristics given by \cite{sweeney2002achieving} often yields effective results. K-anonymization is not a good method to anonymize high-demensional dataset and have performed poorly for certain applications, such as mobile phone datasets \cite{aggarwal2005k}.

\BfPara{$l$-diversity} As an extension to \emph{k}-anonymity, Machanavajjhala \etal proposed \emph{l}-diversity \cite{machanavajjhala2007diversity}, a model that handles some of the weaknesses in the \emph{k}-anonymity model by increasing group diversity for sensitive attributes in the anonymization mechanism. The \emph{l}-diversity model addresses the problem in \emph{k}-anonymity where sensitive attributes within a group exhibit homogeneity. That is, when all values for a sensitive attribute within records in a group are identical, making it easily identifiable. Furthermore, the \emph{l}-diversity is claimed to be resistent against background knowledge attack, whereby an adversary gains information through side-channel means to reduce the set of all possible values for the sensitive attribute. 

\BfPara{Differential Privacy} Dwork introduced the notion of $\varepsilon$-\emph{differential privacy} that provides a definition of privacy in statistical databases. It provides rigorous guarantees against what an adversary can infer from learning the results of some randomized algorithm.  In other words, an algorithm that statisfies differential privacy will not alter the distribution of the output of querying a database regardless whether a particular entry is present or absent from the database. This notion has become an increasingly popular area of research in terms of both theoretical analysis and practical instantiations. Kairouz \etal \cite{kairouz2013composition} have proposed a composition theorem for differential privacy that characterize the level of overall privacy degradation as a function of the number of queries based on hypothesis testing. McSherry and Talwar \cite{mcsherry2007mechanism} have studied the application of differential privacy to digital goods auction where participants are incentivized to be honest. Xiao \etal \cite{xiao2011differential} developed data publishing technique based on wavelet transforms that ensures differential privacy while maintaining accurate answers for range-count queries.

\subsubsection{Redesigning Sharing}
As we highlighted so far, sharing is a complex function, which should take into account various issues, including the structure of the community of trust, the final objective of sharing, and (as advocated in this work) privacy as an additional measure. 

One way to deal with information shared through information sharing schemas and paradigms as private, in its entirety, and provide technical solutions that can address them accordingly by enabling computations on private data. Such computations may include aggregates (private statistics~\cite{kim2012private,rastogi2010differentially,agrawal2000privacy}), set membership (private set operations~\cite{kissner2005privacy,de2010practical}), among others. Techniques for performing such computations may include homorphoic encryption~\cite{brakerski2014efficient,hirt2000efficient,smart2010fully,gentry2009fully,van2010fully} or secure multiple party computation\cite{goldreich1998secure}---among other techniques and variants. 

All of those techniques, when applied to the sheer volume of shared information through the information sharing paradigm, and including simple techniques like $k$-anonymity and $l$-diversity highlighted earlier, yield computationally inefficient solutions. To this end, a practical solution that is aimed towards efficiency should take into account the differential nature of information being consumed through sharing paradigms, as well as the differential nature of community structures in sharing systems. 

One way to achieve efficiency is to consider privacy and risk as highlighted in this paper as an optimization parameter for sharing designs. Taking such parameter into account, it is potentially possible to reduce the size of the data to be processed in a privacy-preserving manner. For example, one could split a given schema of information sharing tool into multiple schemas, based on the classification of the features and attributes of the given sharing standard. As a result, expensive computations that require advanced cryptographic techniques, or are geared towards achieving one of the aforementioned privacy notions---e.g., adding noise for differential privacy, or dummy items to ensure $l$-diversity and $k$-anonymity---could be tolerated on a small subset of attributes of the information being shared, whereas efficient computations could be performed on the raw and plain information that does not have any privacy, sensitive, or confidential markers. 

Privacy is not the only optimization parameter that could be taken into account, but also the structure of the community of trust. For example, highly homogenous and trusted communities, e.g., a result of public-public partnership, could get away without implementing the partitioned architecture for optimization, but rather using minimization (i.e., for what is being shared, and for how long) on the raw data, thus achieving a higher accuracy, and better efficiency. 

Architectural innovation in information sharing is required to improve practicality. Such innovation is facilitated by the differential nature of data and sharing communities, and we argue that they should be taken into to realize efficient sharing solutions. However, to take them into account, further research would be required for understanding the hidden costs in implementing such architecture, the actual trade-off provided by such split architecture, and how to perform complex queries and function (the ultimate purpose of information sharing) on such split architecture, also in a privacy preserving manner.

\section{Quality of Sharing}\label{sec:quality}
So far we have focused on the issue of privacy associated with the sharing, as well as the threat of sharing due to poorly understood communities of trust, which deserve further considerations. Another important research issue in the context of information sharing for actionable intelligence is the quality of shared information. Without high quality of shared information, no actionable intelligence can be obtained. Unfortunately, this issue is not well understood in the literature, and requires further exploration by identifying meaning of quality, and basic methods and tools for assessing it.

We believe that the quality of indicators is of paramount importance to the end-goal of information sharing: a timely indicator, like a source of attack, could be used to defend against an emerging attack, unlike a stale indicators that could be hardly used for postmortem analysis. Thus quality of indicators is a central issue in information sharing, and requires further attention for realizing the proper definitions, tools for quantification, and incentives for improvement. Little work, however, has been done in the literature on understanding this central notion.
 
In section \ref{sec:qp}, we hinted on the potential correlation between the quality of indicators and privacy. However, privacy is not the only factor that affects the quality of indicators (albeit perhaps negatively, when privacy of indicator is ensured). Indicators are often time-sensitive, and time is another way to assess the quality of indicator. Finally, a meaningful annotation and label of the indicator is another potential assessor of the quality of the indicator. 

How to assess the value and quality of an indicator is a nontrivial task: if a consumer in the information sharing community knew the information provided to him through the sharing community, he would not need the sharing of the data in the first place. 

One way to deal with the quality of indicators is to use historical information provided by various community members as a metric for their quality. A community member that provided information that turned to be useful and timely in the past could be annotated as a quality indicators provider, and vice versa. However, such approach for determine the quality of indicators would fall short in multiple aspects. First,  it assesses providers of indicators, rather than individual indicators. Second, certain community members might be well known for certain indicators, e.g., domain names, and other indicators, e.g., binaries, and taking the average of both indicators contributed by them might penalize them, thus not allowing community members to benefit from the (partially) valuable indicators they provide. A possible technique to overcome this shortcoming is to assess community members on various types of indicators, rather than giving them a single score. 

However, even with such consideration, scoring is still for a coarse grained feature of the community member. A scoring that considers history of the community member as a whole, or per class of indicators, and does not assign meaningful scores to individual indicators, is less meaningful. Individual indicators could be more important than the reputation or general quality of the community member, as they are ultimately consumed by the various community members, and their quality indicates the effectiveness of information sharing. 

Recent advances in machine learning and applications to security could be a fruitful direction to answer this question~\cite{JangKWMK15,JangWMYK15,MohaisenA14a,MohaisenA14b,MohaisenWMA14,MohaisenA13} in two ways. First, individual indicators, like a label of malware sample, could be easily vetted through other streams of labels by other vendors~\cite{MohaisenA14b}. Second, labels of indicators, even that are not known to a consumer, could be regenerated by extrapolation, and using machine learning techniques and underlying features of the indicator if the consumer of the indicator had prior samples with the same (or similar label)~\cite{MohaisenA14a}. Exploring how machine learning could be used to assess the quality of indicators is still an open direction that calls for further investigation.

\section{Concluding Remarks}\label{sec:conclusion}
This paper provides a roadmap of issues that need to be explored in order to realize efficient and effective information sharing paradigms for actionable intelligence. With the evolution of the threat and security landscape, no single defender will be able to defend against all threats alone, calling for the utilization of sharing paradigms. However, in order to utilize such paradigms a finer understanding of the various issues associated with sharing is required, including, but not limited to, the underlying community of trust, threat and use models, and privacy highlighted through measurable contexts from various sharing standards and datasets. We argue that utilizing the differential nature of data and communities of trust could be nicely utilized as a feature for optimizing the overhead of sharing, the role that machine learning could play in understanding and assessing the quality of indicators. 

\balance

\end{document}